\documentclass[pra,twocolumn,superscriptaddress,nofootinbib]{revtex4-2}
\usepackage{amsmath}
\usepackage{amsfonts}
\usepackage{amssymb}
\usepackage{mathtools}
\usepackage{graphicx}
\usepackage[caption=false,position=t]{subfig}
\usepackage[colorlinks]{hyperref}
\usepackage[capitalize]{cleveref}
\usepackage{color}

\usepackage{yypreamble}
\newcommand\psubref[1]{\protect\subref{#1}}

\newcommand{\pin}{p_{\rm in}}
\newcommand{\pout}{p_{\rm out}}

\begin{document}

\title{Learning Better Error Correction Codes with Hybrid Quantum-Assisted Machine Learning}

\author{Yariv Yanay}
\email{yanay@umd.edu}
\affiliation{Laboratory for Physical Sciences, 8050 Greenmead Dr., College Park, MD 20740}
\affiliation{Department of Physics, University of Maryland, College Park, MD 20742}


\begin{abstract}
Quantum error correction is one of the fundamental building blocks of digital quantum computation. The Quantum Lego formalism has introduced a systematic way of constructing new stabilizer codes out of basic lego-like building blocks, which in previous work we have used to generate improved error correcting codes via an automated reinforcement learning process. Here, we take this a step further and show the use of a hybrid classical-quantum algorithm. We combine classical reinforcement learning with calls to two commercial quantum devices to search for a stabilizer code to correct errors specific to the device, as well as an induced photon loss error.
\end{abstract}
\maketitle

\section{Introduction}

Quantum error correction is a fundamental requirement for full scale quantum computation \cite{Shor1995}. While recent years have shown remarkable experimental progress in the field \cite{Egan2021a, Ryan-Anderson2021, Bluvstein2024}, 
much of it has focused on well-known schemes such as the surface code \cite{Kitaev2003, Xue2022,Krinner2022,Acharya2025}, while long-term progress may require quantum error correction codes (QECCs) with better scaling in terms of the ratio of physical to logical qubits \cite{Pecorari2025}.
However, the creation of new codes is not straightforward. The Quantum Lego (QL) framework \cite{Cao2022,Farrelly2022,Cao2024, Shen2023} attempts to provide a systematic way of building up new stabilizer codes. QL treats a stabilizer code as a tensor relating the logical state to the physical code words. One can then stitch together multiple codes, via tensor contraction, to generate new ones.

\begin{figure}[htbp] 
   \centering
   \includegraphics[width=\columnwidth]{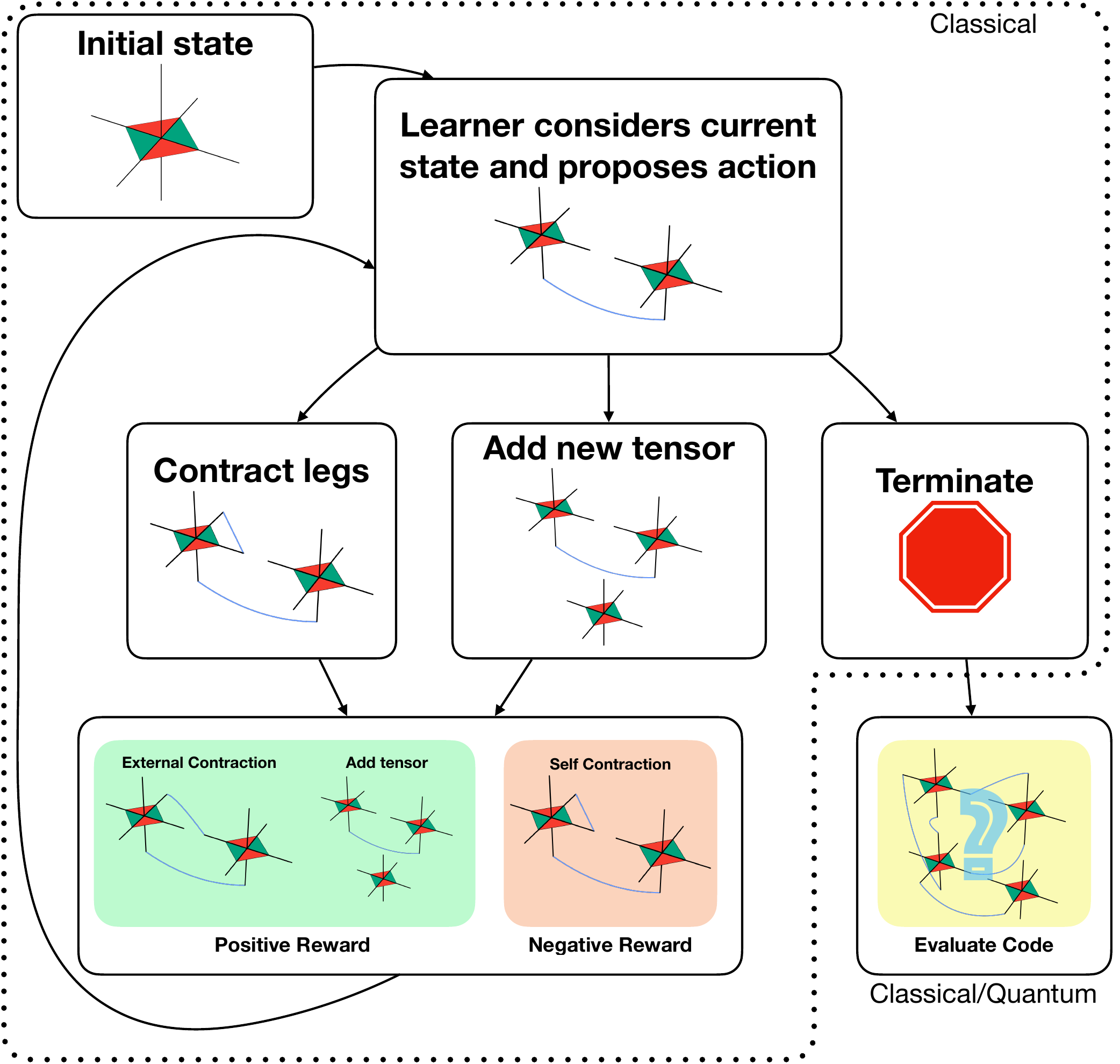} 
   \caption{Reinforcement learning of quantum error correction codes. By using the Quantum Lego framework the learner constructs new codes out of the basic building block tensor. See Ref.~\onlinecite{Su2025} for details. The focus of this work is the final stage, code evaluation. In our previous work this was done classically. Here, we replace this step with calls to quantum computer.}
   \label{fig:sketch}
\end{figure}

While QL systemizes the process of creating new codes, it gives no guarantee of the quality or utility of these codes. In our recent work \cite{Su2025}, we combined that framework with a reinforcement learning approach. By gamifying the process of code concatenation we allow a learning agent to effectively search through the space of possible QECCs, see \cref{fig:sketch}. The choice of a reward function allows us to prioritize different outcomes, such as codes that protect against specific types of biased noise.

Here, we expand the learning agent's reward function into the quantum realm. Evaluating the utility of a QECC is a natural task for a quantum device: by encoding the stabilizer circuit it is easy to calculate its response to noise, either native to the device or injected intentionally. This makes it a natural candidate for a hybrid algorithm \cite{Callison2022}.
We present here demonstrations of both reward functions using two commercially available quantum platforms, Quantinuum's trapped ion processor \cite{quantinuum} and IBM's superconducting chip \cite{ibmquantum,Javadi-Abhari2024}. While we find, predictably, that these platforms are generally above the error-correction threshold for large scale codes, we apply a native noise correction that allows us to see the behavior of a learning process for machines below the threshold.

\section{Summary of Previous Work}

We briefly review the concepts of the code learning process used in our previous work \cite{Su2025}. 

\begin{figure}[tbp] 
   \centering
   \includegraphics[width=\columnwidth]{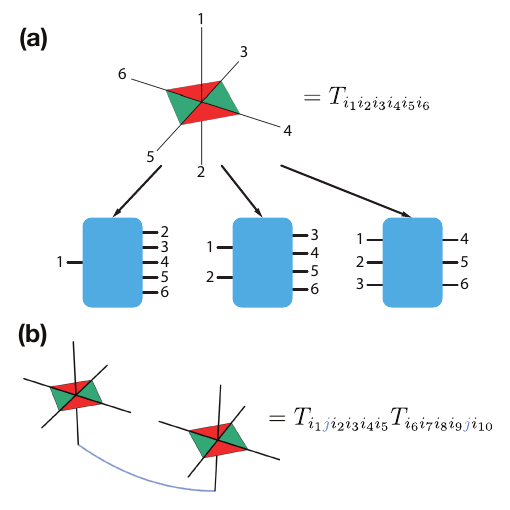}
   \caption{The Quantum Lego formalism. (a) The T6 tensor, pictures, can be any of a variety of QECCs by choosing the logical qubit legs. (b) A new code can be constructed by concatenating two lego blocks, generating a new tensor.}
   \label{fig:legos}
\end{figure}

Our learning routine makes use of the QL formalism to learn new stabilizer codes \cite{Gottesman1997}.
The Quantum Lego framework relies on the identification of tensors with the encoding map of a QECC, $\ket{i_{1},\dotsc,i_{n}} \propto T_{i_{1},\dotsc,i_{n},j_{1},\dotsc,j_{m}}\ket{j_{1},\dotsc,j_{m}}$  for some $m\ge n$, where $\ket{i_{1},\dotsc,i_{n}}$ is the logical state and $\ket{j_{1},\dotsc,j_{m}}$ the corresponding code word. Tensors which obey a required set of symmetries act as the basic lego building blocks. One can build up larger codes by connecting these lego building blocks, corresponding to contracting the indices of two tensors with each other. The combined tensor network inherits all symmetries of the individual legos that satisfy the additional constraints imposed by the fused legs. The transformation of the symmetries under such gluing operations is captured graphically as operator pushing. This is briefly outlined in \cref{fig:legos}. For more information, see Ref.~\onlinecite{Cao2022}.

In Ref.~\onlinecite{Su2025}, we converted the QL formalism into a game, as summarized in \cref{fig:sketch}, and then applied a reinforcement learning routine to search for codes with different reward functions. We used two classical reward functions: we evaluated these codes by calculating code distance, the minimum number of physical Pauli operations needed to go from one code word to a different one, and by the logical error rate, the probability of generating an undetected bit flip. Classically, these are both exponentially difficult, requiring essentially an enumeration of all possible errors up to the code distance. While we are not aware of any quantum algorithm with exponential advantage in calculating code distance, the error rate lends itself to faster calculation via a quantum algorithm.

\section{Hybrid algorithm}

\begin{figure}[tbp] 
   \centering
   \includegraphics[width=\columnwidth]{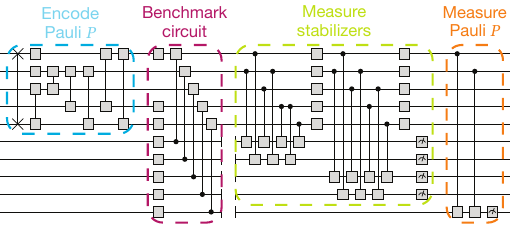} 
   \caption{Quantum evaluation of an error correcting code. We obtain from the classical leaner a stabilizer matrix, which is converted into encoder circuits for $P\in{X,Y,Z}$, generating the state $\ket{\pin}$ (blue). After this, we apply some form of a benchmark circuit (red) that aligns with the expected operation of the qubit. This can include, e.g., logical operations on the qubit, idling for a set amount of time, or artificially inducing errors.
   Then, stabilizers are measured to extract an error syndrome $s$ (yellow), and finally $\pout = \avg{\hat P}$ is measured to observe whether an error has occurred. 
   The set $(\pin,s,\pout)$ is recorded for each run.
   }
   \label{fig:evaluationcirc}
\end{figure}

The uncorrected error rate is straightforward to calculate with a quantum device, and we outline a procedure to do so in \cref{fig:evaluationcirc}. We begin by preparing the qubit is some initial logical state $\ket{\pin}$, where $\pin=\pm P$, $P\in\{X, Y, Z\}$ is some Pauli state; then, we subject the state to the kind of noise we seek to protect against, either simply by allowing for device operation or by intentionally injecting known errors; we apply an error detection cycle, where the stabilizers are measured and recorded; and finally we measure the Pauli $P$ to find $\pout$, comparing it to $\pin$. We then calculate the rate of uncorrected errors.

This form of calculation is generic regardless of the underlying device. However, one must first perform secondary classical calculation to convert the QECC generated by the learning algorithm into a set of quantum algorithms for state encoding, stabilizer measurement, and possibly induced noise. This process is dependent on the particular architecture, its native gates and connectivity constraints.

\begin{figure}[htbp] 
   \centering
   \includegraphics[width=\columnwidth]{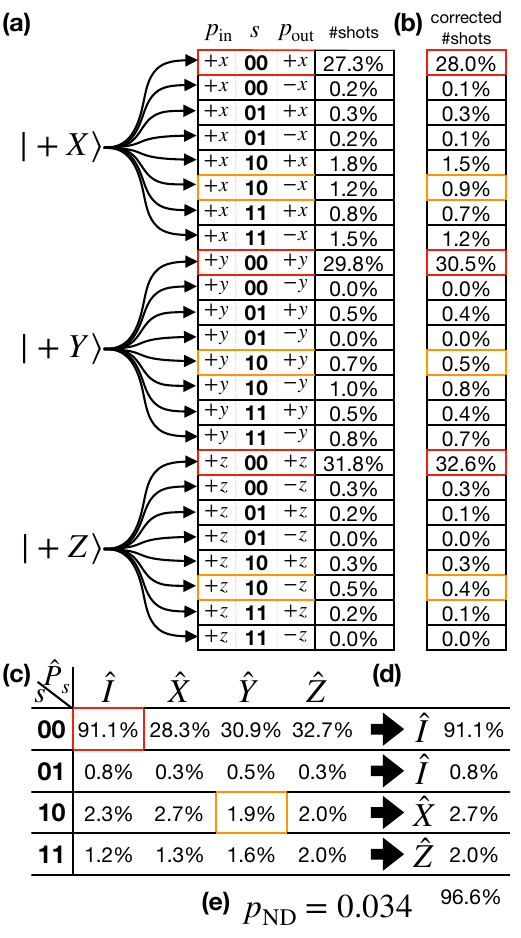} 
	\subfloat{\label{fig:calc-example-res}}
	\subfloat{\label{fig:calc-example-corrected}}
	\subfloat{\label{fig:calc-example-truetable}}
	\subfloat{\label{fig:calc-example-truetot}}
	\subfloat{\label{fig:calc-example-pND}}
   \caption{Sample calculation of the reward function for a simple two-qubit code. 
   \psubref{fig:calc-example-res} Initially we run the code with different initial states, gaining a set of results with syndrome, initial Pauli and error $(\pin,s,\pout)$. 
   \psubref{fig:calc-example-corrected} We correct for the expected machine fidelity, increasing the share of results of $s=0$, $\pout=\pin$ and reducing the rest. 
   \psubref{fig:calc-example-truetable} For each syndrome $s$ and application of a logical Pauli $P_{s}$, we find what portion of the shots would result in the correct logical state, $\hat P_{s}\ket{\pout}=\ket{\pin}$. For example, for $s=00$, $\hat P_{s}=\hat I$, we add up the results from $(+x,00,+x),(+y,00,+y),(+z,00,+z)$ (red highlighting); for $s=10$, $\hat P_{s} =\hat Y$, the logical state of the $\ket{+X},\ket{+Z}$ states would be flipped while the $\ket{+Y}$ would be not, and so we add up $(+x,10,-x),(+y,10,+y),(+z,10,-z)$ (orange highlighting). 
   \psubref{fig:calc-example-truetot} For each syndrome we choose $P_{s}$ to maximize the correct output. 
   \psubref{fig:calc-example-pND} Adding up the portion of correct output, we find the percentage of uncorrected error.  }
   \label{fig:calc-example}
\end{figure}

Note that we do not perform any error correction within the quantum machine. While error detection in a stabilizer code is straightforward, by measurement of the wrong outcome of a syndrome, error correction is generically complex. There is no generic algorithm for deducing an error from a set of stabilizer measurement (or syndrome). Instead, we collect all sets of results, including initial logical state $\pin$, syndrome $s$, and final logical state $\pout$. For each syndrome measured, we assign a logical Pauli correction $\hat P_{s} \in\{\hat I,\hat X,\hat Y,\hat Z\}$, which would have returned the correct final state $\ket{\pin} = \hat P_{s}\ket{\pout}$ for the maximum number of runs. Each run where applying the syndrome Pauli to a final state would not bring it back to the correct initial state is considered uncorrected, and the rate of these to the total number of runs, $p_{\rm ND}$, is used as a cost which we seek to minimize. An example of this process is shown in \cref{fig:calc-example}.

\subsection{Simulation with STIM}
Our first implementation, to demonstrate the feasibility of the learning process, is with the simulation package STIM \cite{Gidney2021}, which allows efficient calculation of Clifford circuits. Our stabilizer circuits are composed entirely of Clifford stabilizers, and so as long as we restrict ourselves to Clifford errors we can simulate the entire hybrid learning process.

\begin{figure}[tbp]
   \centering
   \includegraphics[width=\columnwidth]{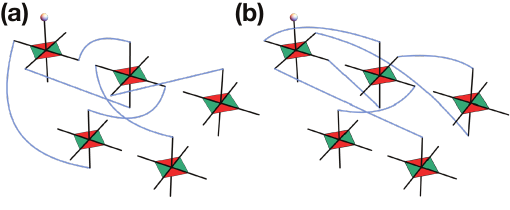} 
	\subfloat{\label{fig:stimcodes-res-nonbias}}
	\subfloat{\label{fig:stimcodes-res-biased}}
   \caption{
   Optimal codes for Pauli noise with \psubref{fig:stimcodes-res-nonbias} isotropic noise, $\Pr(X)=\Pr(Y)=\Pr(Z)=0.01$, and \psubref{fig:stimcodes-res-biased} biased noise $\Pr(X)=\Pr(Y)=0.01, \Pr(Z)=0.05$, discovered by learning using a fast Clifford gate simulator \cite{Gidney2021}. These codes reduce the qubit error by 97\% and 85\%, respectively. }
   \label{fig:stimcodes}
\end{figure}

For STIM, we make use of its native routines for both encoding the state, as well as generating noise. We compare two learning runs, one with isotropic noise, $\Pr(X)=\Pr(Y)=\Pr(Z)=0.01$, and one with biased noise, $\Pr(X)=\Pr(Y)=0.01, \Pr(Z)=0.05$. We show the optimal codes generated by the learning process in \cref{fig:stimcodes}.

\subsection{Natural Errors On Real Devices}

Next, we apply the learning process to two real quantum devices available for academic use. 

First, we use the Quantinuum H1-1 machine, a 20-qubit trapped ion system \cite{quantinuum}. 
Like other trapped ion systems, the H1-1 has all-to-all connectivity between the qubits. This means there is no need to take into account any kind of geometry, and we can use the canonical stabilizers, as supplied by the learning agent. On this machine, to reduce the required quantum calculation time, we use the Quantinuum simulator to evaluate calls for circuits of 10 qubits or less, while larger circuits are evaluated with the real quantum device.

Secondly, we run the learning algorithm on IBM's supercondcucting heavy-hex system \cite{ibmquantum}. The superconducting topology allows only for nearest neighbor two-qubit gates. Longer-ranged operations are broken up into series of multiple gates on adjoining qubits. As such, the position of qubits in a stabilizer code, and the decision of which set of generators of the stabilizer to use, can greatly affect the number of gates and the fidelity of the state encoding and stabilizer measurement operation. 
Thus, for the learning process on the IBM machine, we append a code distance minimization stage to the classical learner. This uses a greedy minimization algorithm to generate a set of stabilizer with the shortest total inter-qubit distance across the stabilizer matrix.

\begin{figure}[t] 
   \centering
   \includegraphics[width=\columnwidth]{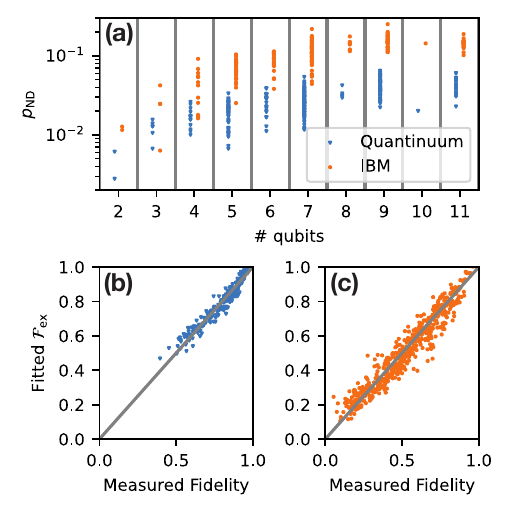} 
	\subfloat{\label{fig:quantum-natural-pND}}
	\subfloat{\label{fig:quantum-natural-F-QTM}}
	\subfloat{\label{fig:quantum-natural-F-IBM}}
   \caption{Results from a learning run with only the natural noise from Quantinuum and IBM machines.
   \psubref{fig:quantum-natural-pND} The calculated $p_{\rm ND}$, fraction of undetected errors, for the learning runs, plotted by the number of qubits in the code. As the noise is above the error correction threshold, the number of errors increases with the number of physical qubits, and the learner ends up minimizing to a two-qubit code.
   To observe the learning process, we find that we can fit the fidelity of these results, see \cref{eq:Ffit}. The results of the fit are shown for \psubref{fig:quantum-natural-F-QTM} Quantinuum and \psubref{fig:quantum-natural-F-IBM} IBM.
   }
   \label{fig:quantum-natural}
\end{figure}

We initially attempt to use the agent to find a stabilizer code suitable for the natural noise of both these devices. To do this, we simply omit the Benchmark Circuit portion of the circuit shown in \cref{fig:evaluationcirc}. The results are shown in \cref{fig:quantum-natural-pND}. We find, not unexpectedly, that the natural error rate of operations on these device is higher than the threshold for any of the codes generated by the QLego learner, and thus the optimal code under the conditions in the minimally achievable one. 

\section{Renormalized Results with Induced Qubit Relaxation}

To obtain a sense of how the learning process would proceed on a higher-fidelity quantum computer, we employ a skewed reward function. From the results generated by the initial learning run, we evaluate the fidelity by fitting 
\begin{equation}
\log\mathcal F_{\rm ex} = c_{q}N_{q} + c_{1}N_{1} + c_{2}N_{2}
\label{eq:Ffit}
\end{equation}
where $N_{q}, N_{1}, N_{2}$ are the number of qubits, one-qubits rotations gates and two-qubit gates in the circuit, respectively. We find that this allows us to correctly predict the fidelity, as shown in \cref{fig:quantum-natural-F-QTM,fig:quantum-natural-F-IBM}.

We thus run a modified learning algorithm on Quantinuum's devices. For each run of the quantum device with number of shots $N_{\rm tot}$, we increase the weight of error free outputs while maintaining the total number of shots, 
\begin{equation}\begin{split}
& N(\pin, s, \pout) \to N(\pin, s, \pout)\times\\ 
& \quad \fopt{ 1/\mathcal (F_{\rm ex})^{\alpha} & s=0, \pout=\pin
 \\ \frac{N_{\rm tot} - N_{0}/(\mathcal F_{\rm ex})^{\alpha}}{N_{\rm tot} - N_{0}} & \text{o/w.} }
\end{split}\end{equation}
where $N_{0} = \sum_{\pin} N(\pin,0,\pin)$.
An example of this process is shown in \cref{fig:calc-example-corrected}.

Next, we introduce artificial error. To demonstrate a quantum advantage calculation, we consider an error that cannot be replicated purely with Clifford gates. One such error typical to many qubit modalities  is relaxation error, where a qubit in the excited (or $\ket{1}$) state relaxes to the ground ($\ket{0}$) state by emission of a photon. The continuous process,
\begin{equation}
\dot{\hat \rho} = \Gamma\p{\hat \gs^{-}\hat\rho\hat \gs^{+} - \half\hat\gs^{+}\hat \gs^{-}\hat\rho - \half\hat \rho\hat\gs^{+}\hat \gs^{-}},
\end{equation}
corresponds to a discrete operation
\begin{equation}\begin{split}
& \hat\rho \to 
\\ & \p{1-\Delta}^{2}\hat\rho + \Delta\p{2-\Delta}\br{\half[1+\hat Z]\hat\rho\half[1+\hat Z] + \hat X\half[1-\hat Z]\hat\rho\half[1-\hat Z]\hat X} 
\\ & +  \frac{\Delta}{2}\p{1-\Delta}\br{\hat\rho - \hat Z\hat \rho\hat Z}
\end{split}\end{equation}
by taking $\Gamma t = 2\log\br{1/\p{1-\Delta}}$. The final term here cannot be generated via a Clifford operation, but the transformation can be achieved using an auxiliary qubit and a partial swap gate,
\begin{equation}
\hat\rho \to \Tr_{a}\br{U_{\ga}\p{\hat\rho\otimes\ket{0}\bra{0}}U_{\ga}\dg}
\end{equation}
with $\hat U_{\ga} = e^{-i\theta\p{\hat X\otimes\hat X + \hat Y\otimes \hat Y}}$, so that $\sin^{2}\theta = \Delta/2$. 

\begin{figure}[t] 
   \centering
   \includegraphics[width=\columnwidth]{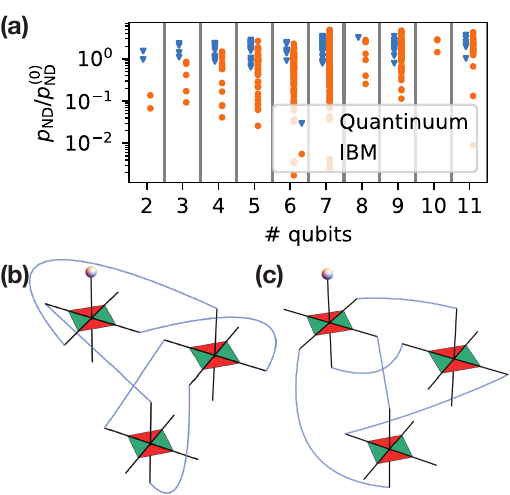} 
	\subfloat{\label{fig:quantum-renormed-circuit}}
	\subfloat{\label{fig:quantum-renormed-qtm}}
	\subfloat{\label{fig:quantum-renormed-ibm}}
   \caption{Results from a learning run renormalized to reduce natural machine noise, with additional qubit relaxation noise added. \psubref{fig:quantum-renormed-circuit} We plot the ratio of the stabilizer code undetected errors to the rate of induced noise, for codes with different number of physical qubits. We find that in both cases the ideal error correction is found for 7-qubit stabilizers. \psubref{fig:quantum-renormed-qtm} The Quantinuum code is renormalized for half its natural infidelity, resulting in a 7-qubit with 45\% improvement, \psubref{fig:quantum-renormed-ibm} The IBM code is renormalized to remove all of its machine infidelity, resulting in a 6-qubit code with 99.8\% improvement.
   }
   \label{fig:quantum-renormed}
\end{figure}

We implement this gate as induced noise for both Quantinuum and IBM machines and run a learning process with renormalized errors to account for internal machine infidelity.
For the Quantinuum machine, we renormalize the results by half the natural noise, choosing $\alpha=0.5$. For the IBM machine, where the nearest-neighbor constraints produce much lower fidelity, we take $\alpha=1$. The results for both runs are shown in \cref{fig:quantum-renormed}. We now observe that the rate of undetected error $p_{\rm ND}$ varies much more with qubit number, and for some codes indeed shows improvement over the bare state. We find an optimal 7-qubit code for the Quantinuum machine, and an optimal 6-qubit code for the IBM machine, both shown in the figure.

\section{Summary}

We're presented here the results of a hybrid learning algorithm, combining the concepts of Quantum Lego and reinforcement learning with calls real quantum devices to test the utility of generated stabilizer codes. 
While current devices are generally not yet at the error correction threshold, we made use of our knowledge of their error profile to renormalize the result, allowing us to demonstrate learning runs that achieve improved fidelities. We've also shown this in the context of qubit relaxation errors do not correspond to classically-simulable Clifford operations.

With better quantum devices available every day, this kind of combination shows the potential of future hybrid algorithm. In particular, this kind of learning would allow moving past human-designed algorithms such as the surface code, and into a variety of codes specialized for device properties of new hardware.

\,
\section*{Acknowledgements}
We would like to thank Matthew Brooks for his assistance in accessing computational resources.

\bibliography{/Users/yarivyanay/Documents/TeX/zotero.bib}

\begin{thebibliography}{10}
\providecommand{\url}[1]{\texttt{#1}}
\providecommand{\urlprefix}{}
\providecommand{\eprint}[2][]{\url{#2}}

\bibitem{Shor1995}
P.~W. Shor, \href{http://dx.doi.org/10.1103/PhysRevA.52.R2493}{Phys. Rev. A
  \textbf {52} R2493,  (1995)}.

\bibitem{Egan2021a}
L.~Egan \emph{et~al.},
  \href{http://dx.doi.org/10.1038/s41586-021-03928-y}{Nature \textbf {598} 281,
   (2021)}.

\bibitem{Ryan-Anderson2021}
C.~{Ryan-Anderson} \emph{et~al.},
  \href{http://dx.doi.org/10.1103/PhysRevX.11.041058}{Phys. Rev. X \textbf {11}
  041058,  (2021)}.

\bibitem{Bluvstein2024}
D.~Bluvstein \emph{et~al.},
  \href{http://dx.doi.org/10.1038/s41586-023-06927-3}{Nature \textbf {626} 58,
  (2024)}.

\bibitem{Kitaev2003}
A.~{\relax Yu}. Kitaev,
  \href{http://dx.doi.org/10.1016/S0003-4916(02)00018-0}{Annals of Physics
  \textbf {303} 2,  (2003)}.

\bibitem{Xue2022}
X.~Xue, M.~Russ, N.~Samkharadze, B.~Undseth, A.~Sammak, G.~Scappucci, and
  L.~M.~K. Vandersypen,
  \href{http://dx.doi.org/10.1038/s41586-021-04273-w}{Nature \textbf {601} 343,
   (2022)}.

\bibitem{Krinner2022}
S.~Krinner \emph{et~al.},
  \href{http://dx.doi.org/10.1038/s41586-022-04566-8}{Nature \textbf {605} 669,
   (2022)}.

\bibitem{Acharya2025}
R.~Acharya \emph{et~al.},
  \href{http://dx.doi.org/10.1038/s41586-024-08449-y}{Nature \textbf {638} 920,
   (2025)}.

\bibitem{Pecorari2025}
L.~Pecorari, S.~Jandura, G.~K. Brennen, and G.~Pupillo,
  \href{http://dx.doi.org/10.1038/s41467-025-56255-5}{Nat Commun \textbf {16}
  1111,  (2025)}.

\bibitem{Cao2022}
C.~Cao and B.~Lackey, \href{http://dx.doi.org/10.1103/PRXQuantum.3.020332}{PRX
  Quantum \textbf {3} 020332,  (2022)}.

\bibitem{Farrelly2022}
T.~Farrelly, D.~K. Tuckett, and T.~M. Stace,
  \href{http://dx.doi.org/10.1088/1367-2630/ac5e87}{New J. Phys. \textbf {24}
  043015,  (2022)}.

\bibitem{Cao2024}
C.~Cao, M.~J. Gullans, B.~Lackey, and Z.~Wang,
  \href{http://dx.doi.org/10.1103/PRXQuantum.5.030313}{PRX Quantum \textbf {5}
  030313,  (2024)}.

\bibitem{Shen2023}
R.~Shen, Y.~Wang, and C.~Cao,
  \href{http://dx.doi.org/10.48550/arXiv.2310.19538}{ arXiv:2310.19538,
  (2023)}.

\bibitem{Su2025}
V.~P. Su, C.~Cao, H.-Y. Hu, Y.~Yanay, C.~Tahan, and B.~Swingle,
  \href{http://dx.doi.org/10.1103/PhysRevApplied.23.034048}{Phys. Rev. Appl.
  \textbf {23} 034048,  (2025)}.

\bibitem{Callison2022}
A.~Callison and N.~Chancellor,
  \href{http://dx.doi.org/10.1103/PhysRevA.106.010101}{Phys. Rev. A \textbf
  {106} 010101,  (2022)}.

\bibitem{quantinuum}
\emph{Quantinuum H1-1}, \urlprefix\url{https://www.quantinuum.com/} (2025).

\bibitem{ibmquantum}
\emph{{{IBM Quantum}}} \url{https://quantum.cloud.ibm.com/}, 2025. In this paper we used \emph{ibm\_kingston}, which is one of the IBM Quantum Heron processors.

\bibitem{Javadi-Abhari2024}
A.~{Javadi-Abhari} \emph{et~al.},
  \href{http://dx.doi.org/10.48550/arXiv.2405.08810}{ arXiv:2405.08810,
  (2024)}.

\bibitem{Gottesman1997}
D.~Gottesman.
\newblock \emph{Stabilizer Codes and Quantum Error Correction}.
\newblock Ph.D. thesis, California Institute of Technology (1997).
\newblock \urlprefix\url{https://arxiv.org/abs/quant-ph/9705052}.

\bibitem{Gidney2021}
C.~Gidney, \href{http://dx.doi.org/10.22331/q-2021-07-06-497}{Quantum \textbf
  {5} 497,  (2021)}.

\end{thebibliography}

\end{document}